\documentclass[aps,amsmath,amssymb,preprint]{revtex4}

\usepackage{graphicx}
\usepackage{dcolumn}
\usepackage{bm}

\begin{document}
\title{Phase Diagram and Quantum phase transition in Newly Discovered Superconductors: $SmO_{1-x}F_xFeAs$ }
\author{R. H. Liu$^1$, G. Wu$^1$, T. Wu$^1$, D. F. Fang$^1$, H. Chen$^1$, S. Y. Li$^2$, K. Liu$^1$,
Y. L. Xie$^1$, X. F. Wang$^1$, R. L. Yang$^1$, L. Ding$^2$, C.
He$^2$, D. L. Feng$^2$ and X. H. Chen$^1$}  \affiliation{1. Hefei
National Laboratory for Physical Science at Microscale and
Department of Physics, University of Science and Technology of
China, Hefei, Anhui 230026, P. R.
China\\
2. Department of Physics and Laboratory of Advanced Materials, Fudan
University, Shanghai 200433, P. R. China }
\date{\today}
\maketitle

{\bf The magnetic fluctuations associated with a quantum critical
point (QCP) are widely believed to cause the non-Fermi liquid
behaviors and unconventional superconductivities, for example, in
heavy fermion systems and high temperature cuprate superconductors.
Recently, superconductivity has been discovered in iron-based
layered compound $LaO_{1-x}F_xFeAs$ with $T_c$=26 K\cite{yoichi},
and it competes with spin-density-wave (SDW) order\cite{dong}.
Neutron diffraction shows a long-rang SDW-type antiferromagnetic
(AF) order at $\sim 134$ K in LaOFeAs\cite{cruz,mcguire}. Therefore,
a possible QCP and its role in this system are of great interests.
Here we report the detailed phase diagram and anomalous transport
properties of the new high-Tc superconductors $SmO_{1-x}F_xFeAs$
discovered by us\cite{chenxh}. It is found that superconductivity
emerges at $x\sim$0.07, and optimal doping takes place in the
$x\sim$0.20 sample with highest $T_c \sim $54 K. While $T_c$
increases monotonically with doping, the SDW order is rapidly
suppressed, suggesting a QCP around $x \sim$0.14. As manifestations,
a linear temperature dependence of the resistivity shows up at high
temperatures in the $x<0.14$ regime, but at low temperatures just
above $T_c$ in the $x>0.14$ regime;  a drop in carrier density
evidenced by a pronounced rise in Hall coefficient are observed,
which mimic the high-$T_c$ cuprates. The simultaneous occurrence of
order, carrier density change and criticality makes a compelling
case for a quantum critical point in this system. }

Since the discovery of high-transition temperature ($T_c$)
superconductivity in layered copper oxides, extensive efforts have
been devoted to explore the higher $T_c$ superconductivity. Very
recently, layered rare-earth metal oxypnictides LnOMPn (Ln=La, Pr,
Ce, Sm; M=Fe, Co, Ni, Ru and Pn=P and As) with ZrCuSiAs type
structure\cite{quebe,zimmer} have attracted great attention due to
the discovery of superconductivity at $T_c=26$ K in the iron-based
$LaO_{1-x}F_x$FeAs (x=0.05-0.12)\cite{yoichi}. Immediately, $T_c$
was drastically raised to 43 K in $SmO_{1-x}F_x$FeAs\cite{chenxh},
followed by reports of $T_c$=41 K in $CeO_{1-x}F_x$FeAs\cite{chen},
and 52 K in $PrO_{1-x}F_x$FeAs\cite{ren}. These discoveries have
generated much interest for exploring novel high temperature
superconductor, and provided a new material base for studying the
origin of high temperature superconductivity.  The superconductivity
in these materials appears to be unconventional, and much careful
work will be required to elucidate the interesting physics here. It
appears that the electron-phonon interaction is not strong enough to
give rise to such high transition temperatures\cite{Boeri}, while
strong ferromagnetic and antiferromagnetic fluctuations have been
proposed to be responsible\cite{cao,dai,ma}.

The undoped material LaOFeAs has been reported to undergo a spin
density wave (SDW) transition at 150 K\cite{yoichi,dong}. The SDW is
suppressed and superconductivity emerges with increasing F
doping\cite{dong}. Systematical characterizations for evolution of
the superconductivity and SDW with F-doping are important for
understanding the underlying physics. Here we successfully prepared
a series of $SmO_{1-x}F_xFeAs$ samples with $x=0 \sim 0.3$ and
systematically studied their resistivity and Hall coefficient, and
gave its phase diagram. The resistivity shows clear anomaly and the
Hall coefficient increases sharply at temperature $T_s \sim 148$ K
for SmOFeAs, indicating the onset of spin-density wave. $T_s$ was
found to decrease with increasing doping, manifesting the
competition between superconductivity and SDW. A crossover occurs
around $x\sim 0.14$ for $T-$linear dependence of resistivity from in
high temperature range to in low temperature range (just above
$T_c$) with increasing doping. The drop in carrier density and a
T-linear dependence of the resistivity are two hallmarks of a
quantum phase transition in metals \cite{lohneysen}, strongly
suggesting existence of a quantum critical point in this system.

Polycrystalline samples with nominal composition
$SmO_{1-x}F_{x}FeAs$ (x=0-0.3) were synthesized by conventional
solid state reaction using high purity SmAs, $SmF_3$, Fe and
$Fe_2O_3$ as starting materials. SmAs was obtained by reacting Sm
chips and As pieces at 600 $^oC$ for 3 hours and then 900 $^oC$ for
5 hours. The raw materials were thoroughly grounded and pressed into
pellets. The pellets were wrapped into Ta foil and sealed in an
evacuated quartz tube. They are then annealed at 1160 $^oC$ for 40
hours. The sample preparation process except for annealing was
carried out in glove box in which high pure argon atmosphere is
filled. Figure 1a shows the XRD patterns for the polycrystalline
samples $SmO_{1-x}F_{x}FeAs$ with different \emph{x}.  The main
peaks in XRD pattern can be well indexed to the tetragonal
ZrCuSiAs-type structure. The XRD patterns show that the samples with
x=0 and 0.05 are single phase. A tiny but noticeable trace of
impurity phases SmOF, Fe and SmAs is observed for $0.12 \leq x
\leq0.20$. Large amount of impurity phases are observed in the
samples with $x>$0.2. Fig.1b shows variation of a-axis and c-axis
lattice parameters with doping F. It shows that both of a-axis and
c-axis lattice parameters decrease systematically with nominal
dopant concentration by substitution of $F^-$ for $O^{2-}$. But the
lattice constants do no change with beyong $x= 0.2$, suggesting that
the chemical phase boundary is reached at $x \sim$0.2 in
$SmO_{1-x}F_{x}FeAs$ system.

Figure 1c shows the superconducting transition of the resistivity
for $SmO_{1-x}F_{x}FeAs$ system. No superconducting transition is
observed down to 5 K for the samples with \emph{x}=0 and 0.05. The
\emph{x}=0.1 sample shows an onset transition at $\sim 12$ K, but no
zero-resistance is observed down to 5 K. The emergence of
superconductivity roughly occurs at $x \sim 0.07$. Superconducting
transition temperature increases monotonically with increasing F
content up to 0.2, optimal doping takes place in \emph{x}=0.2 sample
with highest $T_c \sim $ 54 K as shown in Fig.1d. As shown in
Fig.1a, large amount of impurity phases shows up for the samples
with \emph{x} larger than 0.2, but these samples with \emph{x} up to
0.5 still show superconductivity at $\sim 50$ K, which is nearly
independent of \emph{x}. It further evidences that the F doping is
limited to be about 0.2.

Figure 2 shows the temperature dependence of resistivity in normal
state with temperature up to 475 K for the samples
$SmO_{1-x}F_xFeAs$ (x=0-0.2). The undoped sample shows a similar
behavior to that observed in LaOFeAs\cite{yoichi}. An anomalous peak
associated with SDW shows up at $\sim 150$ K. Below the occurrence
temperature of SDW, the resistivity drops steeply. The temperature
corresponding to the peak in resistivity is defined as the formation
temperature of SDW ( $T_s$ ). F-doping leads to a suppression of the
SDW anomaly peak, and to the shift of $T_s$ to lower temperature. At
10\% F-doping, a weak anomaly peak is still observed. Another
striking feature is that a linear temperature dependence of the
resistivity persists from high temperatures to a characteristic
temperature ($T_0$) in the $x\leq 0.10$ range, at which the
resistivity deviates from T-linear behavior and increases with
decreasing temperature. This should arise from magnetic correlation
before formation of SDW. The $T_0$ of undoped sample is about $\sim
290$ K much higher than $T_s \sim 150$ K of the SDW anomaly peak.
Such a behavior is very similar to that observed in the "stripe"
phase of $La_{1.6-x}Nd_{0.4}Sr_xCuO_4$, where deviation of
resistivity from T-linear behavior also occurs at 150 K above the
"stripe" formation temperature\cite{ishikawa}. In contrast to the
case of the samples with $x \leqslant$ 0.10, no SDW anomaly peak is
observed in the $x\geq0.12$ sample, and the resistivity decreases
more quickly than the linear behavior below $T_0$, being similar to
the pseudogap behavior in high-$T_c$ cuprates\cite{takenaka}. The
steep drop in resistivity below $T_s$ has been ascribed to the
occurrence of SDW \cite{dong}.

Remarkably, the low temperature resistivity can be well fitted with
$a+bT^n$, and the fitting parameter \emph{n} shows a systematical
change from 2.3 to 1 with increasing F content from \emph{x}=0 to
0.15 (Fig.2). An intriguing observation is that a crossover in
temperature dependence of the resistivity happens between the
samples with \emph{x}=0.14 and \emph{x}=0.15. In contrast to the
samples with $x\leq 0.13$, the high-temperature resistivity for the
samples with $x> 0.14$ does not follow a T-linear behavior but tends
to be saturated. However, the temperature dependence of the
low-temperature resistivity just above $T_c$ changes to $T$-linear
dependence, and the resistivity deviates from the T-linear behavior
in high temperatures at certain temperature ( $T_0^\prime $ ).
$T_0^\prime $ increases from 130 K for the \emph{x}=0.15 sample to
205 K for the \emph{x}=0.20 sample. These results indicate that a
profound change in resistivity takes place around
\emph{x}$\sim$0.14, suggesting that the complete suppression of SDW
occurs and quantum critical point appears around \emph{x}=0.14.
Particularly, a possible explanation for T-linear resistivity, that
is widely used to explain the T-linear resistivity in heavy-fermion
metals\cite{lohneysen}, is the scattering of charge carriers by
fluctuation associated with quantum critical point.

Temperature dependence of Hall coefficient ($R_H$) for the
$SmO_{1-x}F_xFeAs$ (x=0-0.2) system with \emph{x}=0, 0.05, 0.10,
0.13, 0.15 and 0.20 is shown in Fig.3a and 3b.  The Hall coefficient
is negative, and decreases with increasing \emph{x}, indicating that
F-doping leads to an increase in carrier concentration. Hall
coefficient for the samples with \emph{x}=0, 0.05 and \emph{x}=0.10
show a strong temperature dependence at low temperatures. As shown
in Fig.3a, Hall coefficient shows a pronounced rise at a certain
temperature which coincides with $T_s$ of the SDW anomaly peaks
observed in Fig.2. It indicates a drop in carrier concentration at
$T_s$ due to the occurrence of SDW. It has been revealed before that
Hall coefficient is prominently enhanced if strong antiferromagnetic
(or SDW) fluctuations exist in heavy fermion system\cite{kontani}.
The evolution of $R_H$ with \emph{x} is very similar to the $R_H$
behavior near AF-QCP in the heavy fermion system $CeMIn_5$
(M=Co,Rh)\cite{nakajima}. Compared to the behavior of \emph{x}=0,
0.05 and 0.10 samples, Hall coefficient of the \emph{x}=0.15 and
0.20 samples shows much weak temperature dependence, and no clearly
pronounced rise at low temperature is observed. Especially, the
\emph{x}=0.2 sample shows very weak temperature dependence at low
temperature. The Hall angle is plotted as $cot \theta _H=\rho/R_H$
vs $T^{1.5}$ in Fig.3c and 3d. It is remarkable that the data make a
straight line in the entire temperature range for the \emph{x}=0.13,
0.15 and 0.20 samples, while in the temperature range above $T_s$
for the \emph{x}=0, 0.05 and 0.10 sample. These results present that
there exists a scaling law between the Hall angle and temperature:
$cot \theta _H \propto T^{1.5}$. The deviation of Hall angle from
$T^{1.5}$ dependence arises from the occurrence of SDW, because the
its characteristic temperature is exactly the same as the
temperature of the SDW anomaly peak in resistivity. Such behavior is
very similar to that observed in high-$T_c$ cuprates where the Hall
angle is proportion to $T^2$\cite{chien}, and deviation of Hall
angle from $T^2$ dependence occurs at the onset temperature of
pseudogap\cite{abe}. The results of resistivity and Hall coefficient
show that a linear temperature dependence of the resistivity and a
drop in carrier density as evidence by a pronounced rise in Hall
coefficient are associated to the occurrence of spin density wave.
Since the T-linear dependence of the resistivity and a drop in
carrier density is the characteristic of a quantum phase transition
in metals\cite{lohneysen}, it suggests a quantum critical point
around \emph{x}=0.14 due to competing of the SDW state and
superconductivity.

Our findings are summarized in the electronic phase diagram shown in
Fig.4, where the characteristic temperatures of $T_s$, $T_0$ and
$T_0^\prime$ are also shown. With increasing F-doping, the onset of
SDW in Sm(O,F)FeAs system is driven down in temperature, and the
superconducting state emerges at $x\sim 0.07$, reaching a maximum
$T_c$ of $\sim 54$ K at \emph{x}=0.20. Compared to the phase diagram
of high-$T_c$ cuprates, no decrease of $T_c$
 is observed with increasing doping up to the chemical phase
boundary at $x\sim 0.20$. Moreover, there is a large intermediate
regime where superconductivity and SDW coexist for Sm(O,F)FeAs.
Based on the evolutions of the resistivity with $T$ and \emph{x} in
Fig.2, the different power law dependent behaviors of the
resistivity are summarized in Fig.4. One could find drastically
different temperature-dependencies of resistivity at two sides of
\emph{x}=0.14. On the left side, the \emph{n}, as that used in the
$a+bT^n$ to fit the low temperature resistivity, continuously
decreases from 2.3 to 1.2 when $x$ is raised from 0 to 0.14
respectively; while on the right side, the \emph{n} is fixed at 1
for $x>0.15$. It clearly indicates that  SDW eventually disappears
around $x \sim 0.14$. More importantly, together with the results of
Hall coefficient, it suggests the existence of a SDW quantum
critical point, which may be crucial for the mechanism of
superconductivity in these iron-based high-$T_c$ superconductors, as
being suggested before for the superconductivity in the copper-based
ones.

\vspace*{2mm} {\bf Acknowledgment:} This work is supported by the
Nature Science Foundation of China and by the Ministry of Science
and Technology of China and STCSM

\vspace*{2mm} {\bf Author information:} The authors declare no
competing financial interests. Correspondence and requests for
materials should be addressed to X. H. Chen (chenxh@ustc.edu.cn).

\newpage
\begin{figure}
\includegraphics[width=\textwidth]{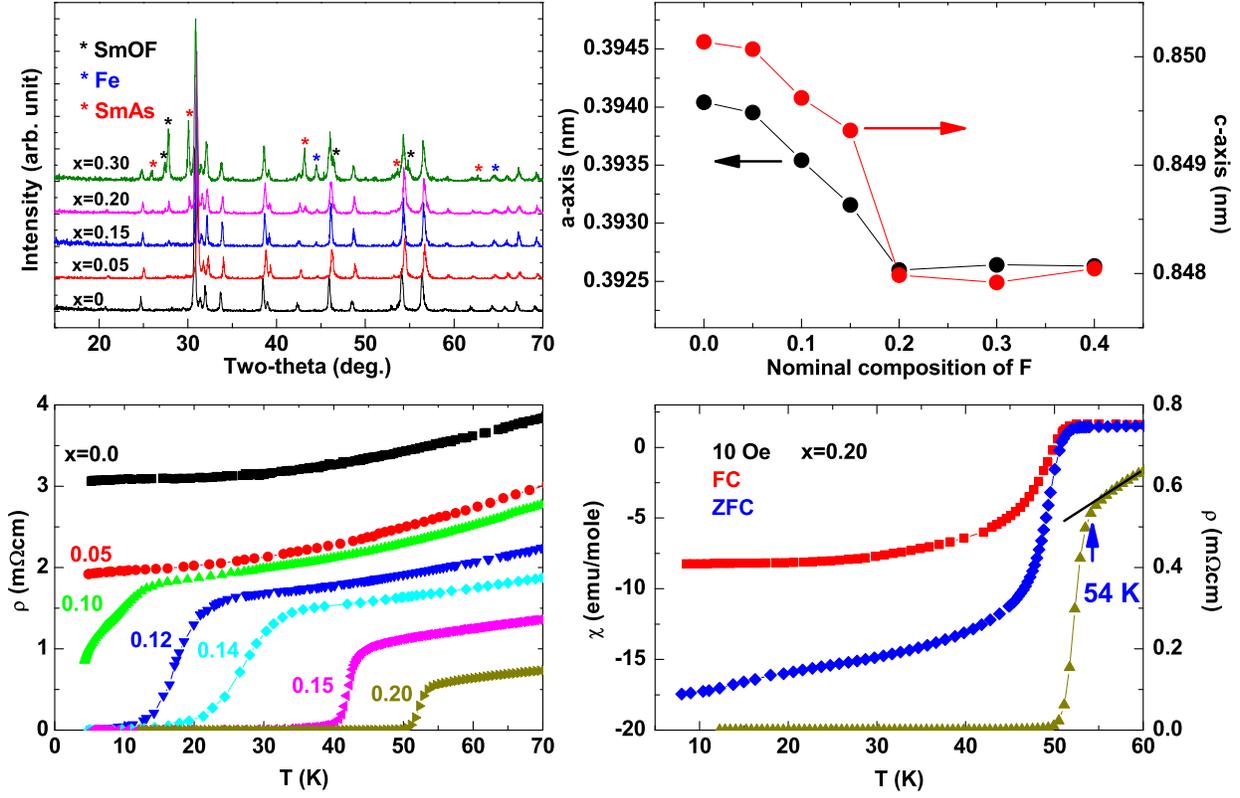}
\caption{(a): X-ray diffraction patterns at room temperature for the
samples $SmO_{1-x}F_xFeAs$. (b): Variation of lattice parameters
with \emph{x}. (c): Superconducting transitions for the samples
$SmO_{1-x}F_xFeAs$ with \emph{x}=0, 0.05, 0.10, 0.12, 0.14, 0.15 and
0.20. (d): Superconducting transition in resistivity and
susceptibility for the \emph{x}=0.20 sample with highest $T_c \sim$
54 K.
\\}
\end{figure}

\newpage
\begin{figure}
\includegraphics[width=0.8\textwidth]{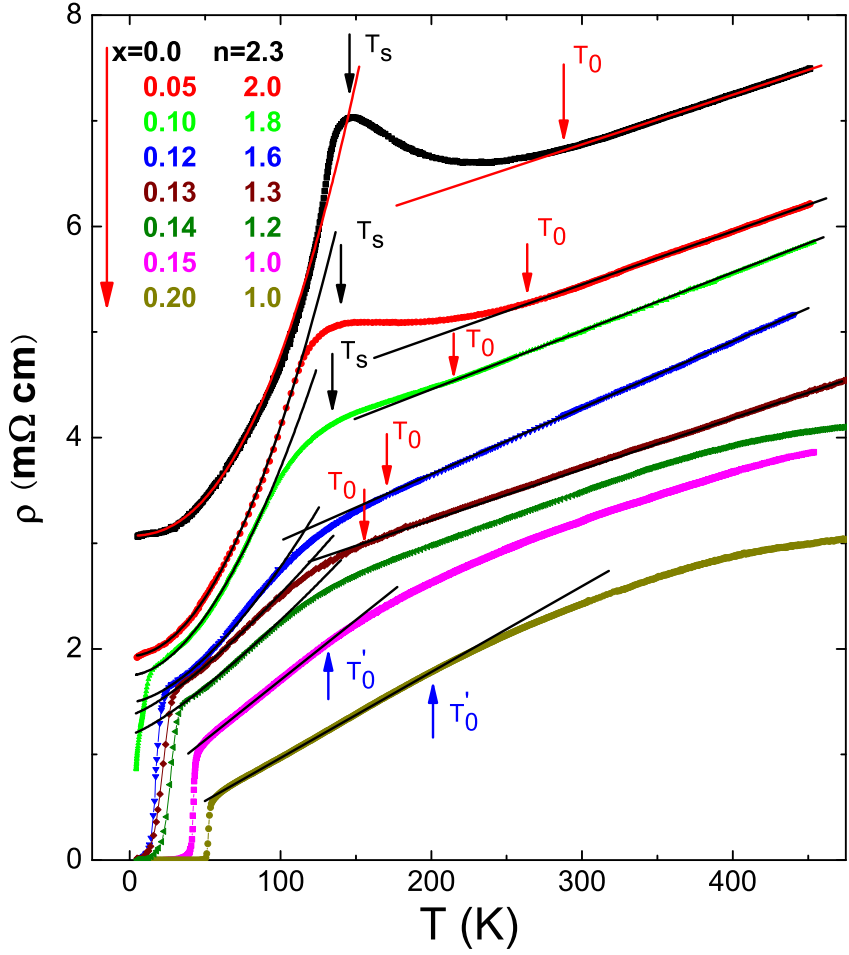}
\caption{ Temperature dependence of the resistivity in the normal
state up to 475 K for the samples $SmO_{1-x}F_xFeAs$ with
\emph{x}=0, 0.05, 0.10, 0.12, 0.13, 0.14, 0.15 and 0.20. Arrows
denote the onset temperature ($T_s$) of spin-density wave and
deviating temperatures from T-linear behavior ($T_0$ and
$T_0^\prime$). The resistivity shows a linear temperature dependence
above $T_0$ for the samples with $x<0.14$, while a T-linear behavior
is observed in the temperature range from $T_c$ to $T_0^\prime$ for
the samples with $x>0.14$. The low temperature resistivity can be
well fitted with $a+bT^n$ for all the samples with $x\leq 0.14$, and
the n=2.3, 2.0, 1.8, 1.6, 1.3, 1.2 for the samples with \emph{x}=0,
0.05, 0.10, 0.12, 1.3, and 1.4, respectively; while low temperature
resistivity shows a T-linear behavior for the samples with
$x\geq0.15$.
\\}
\end{figure}

\newpage
\begin{figure}
\includegraphics[width=\textwidth]{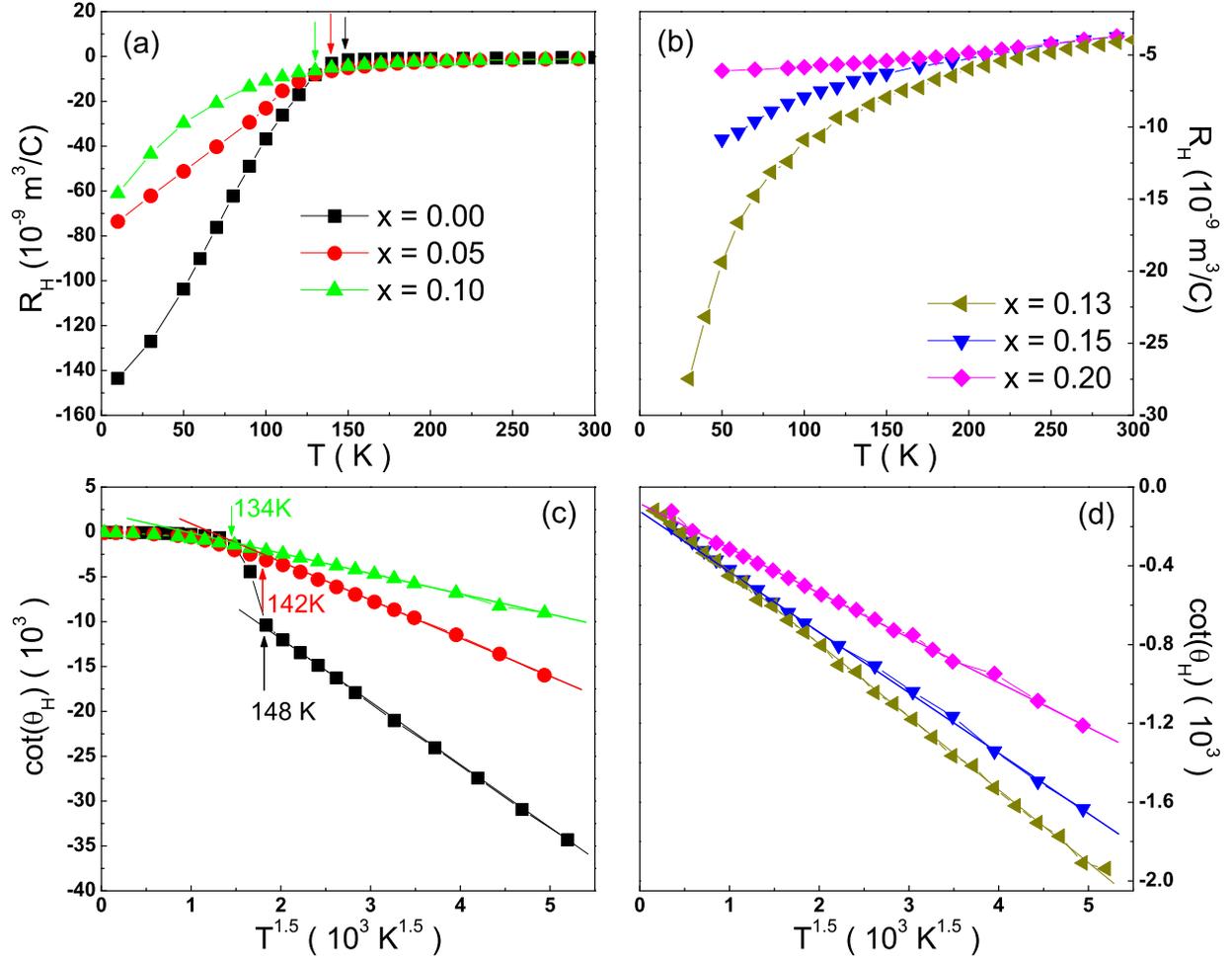}
\caption{ (a): Temperature dependence of Hall coefficient for the
samples $SmO_{1-x}F_xFeAs$ with x=0, 0.05 and 0.10; (b): Temperature
dependence of Hall coefficient for the samples with \emph{x}=0.13,
0.15 and 0.20. A pronounced rise in Hall coefficient is observed for
the \emph{x}=0, 0.05 and 0.10 samples. (c): Hall angle is plotted as
$cot \theta _H$ vs $T^{1.5}$ for the samples with \emph{x}=0, 0.05
and 0.10; (d): $cot \theta _H$ vs $T^{1.5}$ for the samples with
\emph{x}=0.13, 0.15 and 0.20. The data points fall on a straight
line in the entire temperature range for the \emph{x}=0.13, 0.15 and
0.20 samples, while deviation from the straight line at certain
temperature for the \emph{x}=0, 0.05 and 0.10 sample. The deviating
temperature from straight line is exactly the same as $T_s$ of SDW
anomaly peak observed in Fig.2.
\\}
\end{figure}

\newpage
\begin{figure}
\includegraphics[width=\textwidth]{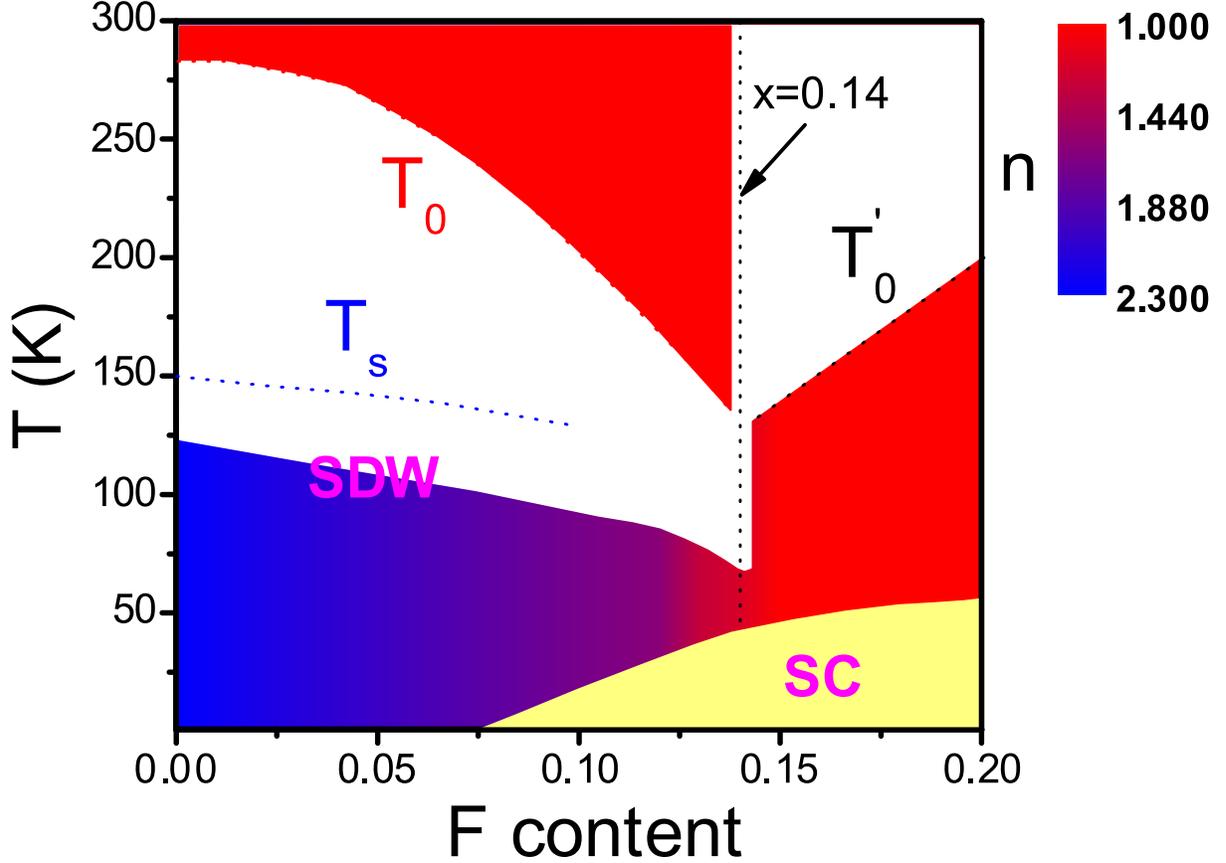}
\caption{ Electronic phase diagram for $SmO_{1-x}F_xFeAs$ system.
$T_s$ indicates the formation temperature of SDW. $T_0$ and
$T_0^\prime$ represent the deviating temperature from a T-linear
dependence of resistivity in low and high temperatures,
respectively. The superconductivity starts to appear at $x\sim
0.07$, reaching a maximum $T_c$ of $\sim$ 54 K at \emph{x}=0.20.
Compared to the phase diagram of high-$T_c$ cuprates, no decrease of
$T_c$ with increasing doping has been observed. It is because the
chemical phase boundary is reached at $x\sim 0.20$. The different
color regions represent different n in the formula $\rho=a+bT^n$,
which can be well used to fitted to the low temperature resistivity
shown in Fig.2. The dot line of \emph{x}=0.14 clearly shows a
boundary for different behavior of T-dependent resistivity below and
above \emph{x}=0.14, suggesting a quantum critical point around
$x\sim 0.14$.
\\}
\end{figure}

\clearpage
 \setcounter{figure}{0}

\renewcommand{\thefigure}{\arabic{figure}s}

\end{document}